\documentclass[11pt,fleqn,twoside]{article}
\usepackage{amsfonts,amssymb,latexsym}
\makeatletter
\newcommand{\prava}[1]{\small\it
\begin{flushleft}
Copyright \copyright \ 2000 by  #1
\end{flushleft}}

\newcommand{\name}[1]{\begin{flushleft}
                       \LARGE \bf #1
                       \end{flushleft}\vspace{-3mm}}

\newcommand{\Author}[1]{\begin{flushleft}
                       \it #1 \end{flushleft}}

\newcommand{\Date}[1]{\begin{flushleft}
                      \small  \it #1 \end{flushleft}}

\newcommand{\ehkol}{Author \ name}
\newcommand{\ohkol}{Article \ name}
\renewcommand{\@evenhead}{
\hspace*{-3pt}\raisebox{-15pt}[\headheight][0pt]{\vbox{\hbox to \textwidth 
{\thepage \hfil \ehkol}\vskip4pt \hrule}}}
\renewcommand{\@oddhead}{
\hspace*{-3pt}\raisebox{-15pt}[\headheight][0pt]{\vbox{\hbox to \textwidth 
{\ohkol \hfil \thepage}\vskip4pt\hrule}}}
\renewcommand{\@evenfoot}{}
\renewcommand{\@oddfoot}{}

\newcommand{\be}{\begin{equation}}
\newcommand{\ee}{\end{equation}}
\newcommand{\ba}{\hspace*{-5pt}\begin{array}}
\newcommand{\ea}{\end{array}}

\makeatother
\long\def\@makecaption#1#2{\vskip 10\p@ \hbox to\hsize{\hfil#1\hfil}}
\newbox\ALH
\setbox\ALH=\hbox{\unitlength=1mm
 \begin{picture}(2.40,2.00)
 \put(1.20,1.00){\circle{2.00}}
 \put(1.20,1.00){\makebox(0,0)[cc]{$\cdot$}}
 \end{picture}}
\newbox\ALHindex
\setbox\ALHindex=\hbox{\unitlength=1mm
 \begin{picture}(1.68,1.40)
 \put(0.84,0.70){\circle{1.40}}
 \put(0.84,0.70){\makebox(0,0)[cc]{$\cdot$}}
 \end{picture}}
\def\alh{\mathchoice{\copy\ALH}{\copy\ALH}{\copy\ALHindex}{\copy\ALHindex}}

\begin{document}
\thispagestyle{empty}
\setcounter{page}{94}

\renewcommand{\ehkol}{I.G. Korepanov}
\renewcommand{\ohkol}{Particles and Strings in a $2+1$-D 
Integrable Quantum Model} 

\begin{flushleft}
\footnotesize \sf
Journal of Nonlinear Mathematical Physics \qquad 2000, V.7, N~1,
\pageref{korepanov-fp}--\pageref{korepanov-lp}.
\hfill {\sc Review Article}
\end{flushleft}

\vspace{-5mm}

\renewcommand{\footnoterule}{}
{\renewcommand{\thefootnote}{} 
 \footnotetext{\prava{I.G. Korepanov}}}

\name{Particles and Strings in a \\{\mathversion{bold}$2+1$}-D 
Integrable Quantum Model}\label{korepanov-fp}

\Author{I.G. KOREPANOV~\footnote{Permanent Address:
South Ural State University, 76 Lenin av., Chelyabinsk 454080,
Russia}

\vskip 2mm
Ural Division of International Institute for Nonlinear Research, Ufa,
Russia\\
E-mail: igor@prima.tu-chel.ac.ru }

\Date{Received February 22, 1999; Revised July 22, 1999; Accepted
September 11, 1999}

\begin{abstract}
\noindent
We give a review of some recent work on generalization of the Bethe
ansatz in the case of $2+1$-dimensional models of quantum f\/ield theory.
As such a model, we consider one associated with the tetrahedron
equation, i.e. the $2+1$-dimensional generalization of the famous
Yang--Baxter
equation. We construct some eigenstates of the transfer matrix of that
model. There arise, together with states composed of point-like particles
analogous to those in the usual $1+1$-dimensional Bethe ansatz,
new string-like states and string-particle hybrids.
\end{abstract}

\renewcommand{\theequation}{\thesection.\arabic{equation}}
\setcounter{equation}{0}

\section*{Introduction}

The Hamiltonians of $1+1$-dimensional ($1+1$-D) integrable quantum models,
i.e. mo\-dels with one-dimensional space and one-dimensional time, are
diagonalized by means of the {\em Bethe ansatz}~\cite{bethe}.
This means that their eigenvectors are searched for in the special form,
namely as a superposition of several pointlike particles with def\/inite
momenta or ``quasimomenta''. The same Bethe ansatz is used for
diagonalizing transfer matrices of two-dimensional statistical physics
models. Clearly, the problem of extending the Bethe ansatz onto the
$2+1$-D models, or maybe f\/inding some other method for constructing their
eigenvectors, is of great importance. In this paper, we are going to present
some results obtained in this direction.

Let us recall the main features of the usual Bethe ansatz. We will have
in mind a model with {\em discrete\/} space, like $XXZ$ spin chain.
In the usual Bethe ansatz, it is supposed that in the vector space
of system states there are naturally distinguished ``0-particle'',
``1-particle'', etc. subspaces that are eigensubspaces of the
Hamiltonian. As a rule, the 0-particle subspace is simply one-dimensional.
Diagonalizing the Hamiltonian in the 1-particle particle space also is
not hard work --- it is suf\/f\/icient to diagonalize the ``quasimomentum''
operator (commuting with the
Hamiltonian), i.e.  the operator of translation
by one lattice unit, at least if we are considering a homogeneous
f\/inite chain with periodic boundary conditions or homogeneous
inf\/inite chain.
In the latter case, we can obtain {\em formal\/} eigenstates if we allow
for any quasimomentum eigenvalues, or {\em generalized\/} eigenstates
(in the sense of functional analysis) if we require the module of
eigenvalues be unity.

A remarkable property of integrable models is that we can construct a
superposition of $n$ one-particle states, lying in the $n$-particle subspace
and being an eigenvector of the Hamiltonian. In the case of a f\/inite
closed chain, there arise severe constraints on quasimomenta --- {\em Bethe
ansatz equations}. The essence of those equations is that the ``amplitude''
corresponding to a given particle must not change after its having made
a complete revolution around the chain, undergoing in its way
the ``scattering'' on all other particles.

Algebraically, the problem is much easier for the inf\/inite chain,
because in such case revolutions around the chain don't exist.
The conditions that single out generalized eigenvectors from the formal
eigenvectors are also quite evident: the ``amplitude'' must not grow
exponentially when the coordinate of any particle tends to the left
or right inf\/inity.

Now let us pass to the $2+1$-D models. Integrable $2+1$-D models are those
related to the {\em tetrahedron
equation\/} whose f\/irst solutions were found by
Zamolodchikov~\cite{zamolodchikov}. The f\/irst solutions for the ``vertex''
modif\/ication of tetrahedron equation were found in paper~\cite{k1}
using some new structure called {\em tetrahedral Zamolodchikov
algebra\/}~\cite{k-trig,k-alg-geo,k2}. As a further development,
Zamolodchikov's solutions were generalized in papers~\cite{bazhanov baxter
1,bazhanov baxter 2,mang kash strog 1,mang kash strog 2}
by Bazhanov, Baxter, Mangazeev, Kashaev, and Stroganov. The discovery of the
interrelation between original Zamolodchikov model and the vertex model of
paper~\cite{k1} is due to Sergeev, Boos, Mangazeev, and
Stroganov~\cite{chetvero}. Before that, J.~Hietarinta~\cite{h} studied
general possibilities for writing down modif\/ications of tetrahedron equation,
and found some new, with respect to paper~\cite{k1}, solutions of its
vertex form.

An attempt to generalize the Bethe ansatz for such
models runs at once into dif\/f\/iculties, because the known solutions of
tetrahedron equation (out of which one can e.g. build a transfer matrix and
search for its eigenvalues) do not allow to single out even
a one-dimensional space. Certainly, in the $1+1$-D case there may also
occur more complicated models than $XXZ$ for which the above stated scheme
of Bethe ansatz requires considerable modif\/ications, but even the experience
of studying those more complicated
models did not yet yield any indications as to how to act
in $2+1$ dimensions. Moreover, the experience of studying {\em classical\/}
integrable equations like Kadomtsev--Petviashvili suggests that in $2+1$
dimensions we should expect greater variety of states than simply
collections of pointlike particles.

In this paper, we will bring together the results of f\/ive
preprints~\cite{5 preprintov} by the author.
We will present some eigenvectors for the transfer matrix
corresponding to the simplest (and historically f\/irst~\cite{k1})
nontrivial solution of the ``vertex type'' tetrahedron equation.
They will be basically formal eigenvectors for the inf\/inite lattice
obeying, however, some conditions of ``good behaviour'' at the inf\/inity.
We will also present some eigenvectors for the f\/inite closed lattice
(``lattice on the torus'').

Our starting point will be the fact that although we could not succeed
in f\/inding what might be called a one-dimensional space invariant under
the action of transfer matrix, there exists a one-particle space whose
{\em subspace is invariant\/} under it. This will allow us to construct
{\em some\/} one-particle and then two-particle states.

The other starting point will be string-like states that arise naturally
 from the {\em tetrahedral Zamolodchikov algebra} --- an algebraic structure
using which the f\/irst ever solutions of the vertex type tetrahedron equation
have been constructed. With all their simplicity, such strings
are a new phenomenon specif\/ic for multidimensional models.

The culmination of this paper is achieved in uniting the particles
and the strings. Here more general solutions of the tetrahedron equation
due to Sergeev, Mangazeev and Stroganov~\cite{mss} will be of use.
We will provide examples showing how one can, with the help of strings,
remove the obstacles on the way of constructing multi-particle states.

Before explaining the details, let us mention here some other interesting
works whose relation with the subject of this paper is not yet completely
clear. In paper~\cite{boos}, H.~Boos studied some ``strings'' in a
two-dimensional $XY$ spin lattice. In papers~\cite{bm1,bm2}, Boos and
Mangazeev obtain functional relations for the eigenvalues of a transfer
matrix consisting of {\em three\/} plane layers.

Our model will be a model on the cubic lattice: the lattice
vertices are points whose all three coordinates are integers.
In each vertex there will be an ``$S$-operator'' (or ``$R$-operator''
in the notations of~\cite{mss}. In the present paper, we are using
the letter $R$ for other purposes) acting in the tensor
product of three linear spaces attached to the {\em links}.
The transfer matrix we will be dealing with will be a ``diagonal'' one:
it is cut out of the lattice by two planes perpendicular to the vector
$(1,1,1)$ in such a way that it consist of separate, not linked to
each other, vertices that can be imagined as placed on one of those
planes like anti-tank hedgehogs. In each of the
planes, the intersection with the cubic lattice yields
a kagome lattice consisting, as is known, of triangles and hexagons.
So it is not surprising that we will use as an important auxiliary tool
one more transfer matrix, of the kagome form, whose image can be
found in this paper as Fig.~\ref{figII-1}.

The $S$-operators are obtained as follows~\cite{k-alg-geo,k2}.
We start from the usual ($1+1$-D) $L$-operators
\[
L=\pmatrix{a&&&d\cr &b&c&\cr &c&b&\cr d&&&a}
\]
obeying the ``free-fermion condition''
\[
a^2+b^2=c^2+d^2.
\]
If $L$ and $M$ are two such operators with the same ratio $cd/ab$,
then the usual Yang--Baxter equation
\be
R^0LM=MLR^0
\label{vvedenie eq *}
\ee
holds with some operator $R^0$ of the same kind.

It is remarkable that, besides (\ref{vvedenie eq *}), one more similar
relation holds:
\[
(R^1)^{\rm T}LM=MLR^1,
\]
where the superscript $\rm T$ means matrix transposing and $R^1\ne R^0$.

Each of the operators $R^0$ and $R^1$ acts in the tensor product of two
two-dimensional vector spaces, and thus can be represented by a matrix
with two upper and two lower indices, or graphically --- as a vertex
where four edges meet. We now combine $R^0$ and $R^1$ in one
object $R^a$, $a=0,1$, with one more edge (``f\/ifth leg'') bearing the
index~$a$.

Some special triple of $R$'s enters in the following {\em defining relation
of the tetrahedral Zamolodchikov algebra}, where we use
the subscripts of $R$ just to indicate the numbers
of spaces where this $R$ acts nontrivially:
\be
R_{12}^a \tilde R_{13}^b \skew5 \tilde{\tilde R}\vphantom R_{23}^c =
 \sum_{d,e,f} S_{def}^{abc}
 \skew5 \tilde{\tilde R}\vphantom R_{23}^f \tilde R_{13}^e R_{12}^d.
\label{tza}
\ee
This is illustrated by Fig.~\ref{figIII-1}.
\begin{figure}[!ht]
\begin{center}
\unitlength=1.00mm
\linethickness{0.4pt}
\begin{picture}(134.00,63.00)
\put(77.00,8.00){\line(4,1){56.00}}
\put(86.00,33.00){\line(3,-1){48.00}}
\put(80.00,0.00){\line(1,2){20.00}}
\put(85.00,10.00){\line(1,4){13.00}}
\put(125.00,20.00){\line(-1,1){38.00}}
\put(95.00,30.00){\line(0,1){33.00}}
\put(85.00,10.00){\circle*{2.00}}
\put(95.00,30.00){\circle*{2.00}}
\put(125.00,20.00){\circle*{2.00}}
\put(95.00,50.00){\circle*{2.00}}
\put(2.00,18.00){\line(4,1){56.00}}
\put(1.00,23.00){\line(3,-1){48.00}}
\put(35.00,0.00){\line(1,2){20.00}}
\put(50.00,30.00){\line(1,4){6.00}}
\put(10.00,20.00){\line(-1,1){10.00}}
\put(40.00,10.00){\line(0,1){35.00}}
\put(0.00,31.00){\makebox(0,0)[lb]{$a$}}
\put(39.00,45.00){\makebox(0,0)[rb]{$b$}}
\put(55.00,54.00){\makebox(0,0)[rb]{$c$}}
\put(87.00,57.00){\makebox(0,0)[rt]{$a$}}
\put(94.00,63.00){\makebox(0,0)[rc]{$b$}}
\put(99.00,62.00){\makebox(0,0)[lc]{$c$}}
\put(0.00,23.00){\makebox(0,0)[rc]{1}}
\put(2.00,17.00){\makebox(0,0)[ct]{2}}
\put(34.00,1.00){\makebox(0,0)[rb]{3}}
\put(85.00,33.00){\makebox(0,0)[rc]{1}}
\put(77.00,9.00){\makebox(0,0)[cb]{2}}
\put(79.00,0.00){\makebox(0,0)[rb]{3}}
\put(10.00,20.00){\circle*{2.00}}
\put(40.00,10.00){\circle*{2.00}}
\put(50.00,30.00){\circle*{2.00}}
\put(66.00,22.00){\makebox(0,0)[cc]{\Large$=$}}
\put(109.00,37.00){\makebox(0,0)[lb]{$d$}}
\put(96.00,42.00){\makebox(0,0)[lc]{$e$}}
\put(87.00,22.00){\makebox(0,0)[rb]{$f$}}
\put(10.00,22.00){\makebox(0,0)[lb]{$R$}}
\put(40.00,8.00){\makebox(0,0)[lt]{$\tilde R$}}
\put(50.00,29.00){\makebox(0,0)[lt]{$\skew4 \tilde{\tilde R}$}}
\put(97.00,51.00){\makebox(0,0)[lc]{$S$}}
\put(85.00,9.00){\makebox(0,0)[lt]{$\skew4 \tilde{\tilde R}$}}
\put(95.00,28.00){\makebox(0,0)[lt]{$\tilde R$}}
\put(125.00,22.00){\makebox(0,0)[lb]{$R$}}
\end{picture}
\end{center}

\vspace{-5mm}

\caption{}
\label{figIII-1}
\end{figure}

Matrix $S$ is exactly the matrix satisfying the vertex-type tetrahedron
equation. It depends on (the dif\/ferences of) three parameters, say
$\varphi_1$, $\varphi_2$ and $\varphi_3$, and its matrix elements
are~\cite{k2}:
\be
S_{000}^{000}=S_{011}^{011}=S_{101}^{101}=S_{110}^{110}=1,
\label{matrica S 1}
\ee
\be
S_{010}^{001}=S_{001}^{010}=-S_{111}^{100}=-S_{100}^{111}=
\root\of{\coth(\varphi_1-\varphi_3)}
\,\root\of{\tanh(\varphi_2-\varphi_3)},
\ee
\be
S_{100}^{001}=S_{111}^{010}=-S_{001}^{100}=-S_{010}^{111} =
\root\of{\tanh(\varphi_1-\varphi_2)}
\,\root\of{\tanh(\varphi_2-\varphi_3)},
\ee
\be
S_{111}^{001}=S_{100}^{010}=S_{010}^{100}=S_{001}^{111}=
\root\of{\tanh(\varphi_1-\varphi_2)}
\,\root\of{\coth(\varphi_1-\varphi_3)},
\label{matrica S 4}
\ee
the other elements are zeroes, and the tetrahedron equation itself is
\be
\ba{l}
S_{01,02,12}(\varphi_0,\varphi_1,\varphi_2)
S_{01,03,13}(\varphi_0,\varphi_1,\varphi_3)
S_{02,03,23}(\varphi_0,\varphi_2,\varphi_3)
S_{12,13,23}(\varphi_1,\varphi_2,\varphi_3)
\vspace{2mm}\\
 =
S_{12,13,23}(\varphi_1,\varphi_2,\varphi_3)
S_{02,03,23}(\varphi_0,\varphi_2,\varphi_3)
S_{01,03,13}(\varphi_0,\varphi_1,\varphi_3)
S_{01,02,12}(\varphi_0,\varphi_1,\varphi_2).
\ea \label{tetr}\hspace*{-8.4pt}
\ee
Here again the subscripts indicate the numbers of spaces where an
operator acts nontrivially, e.g. in
$S_{12,13,23}=(S_{def}^{abc})_{12,13,23}^{\vphantom{hm}}$
the indices $a$ and $d$ correspond to the space number 12, the indices
$b$ and $e$ --- to the space number 13, and the indices $c$ and $f$ --- to
the space number 23.

Note that in Section~\ref{IV}
we will be using more general solutions of the tetrahedron equation
due to Sergeev, Mangazeev and Stroganov~\cite{mss}
together with their parameterization dif\/ferent from
(\ref{matrica S 1})--(\ref{matrica S 4}).

The contents of the remaining sections is as follows.
In Sections \ref{I} and \ref{II} we construct some particle-like states;
in Section \ref{III} we construct the simplest strings;
in Section \ref{IV} we construct a hybrid of a string and a particle;
and in Section \ref{V} we construct hybrids of several strings and
particles. A Discussion completes the paper.

\setcounter{equation}{0}

\section{One- and two-particle states}
\label{I}

As stated in the Introduction, our ``hedgehog'' transfer matrix --- let us
denote it $T$ --- is cut out of the cubic lattice by two planes
perpendicular to the vector $(1,1,1)$. In each of those
planes, the intersection with the cubic lattice yields
a kagome lattice consisting, as known, of triangles and hexagons.
We can group all vertices of the kagome lattice in triples --- vertices
of triangles --- in such a way that
the transfer matrix acts on each triangle separately, turning it
inside out and making a linear transformation in the tensor product
of three corresponding subspaces.

Consider one of such triangles.
The tensor product of three subspaces corresponding to its vertices
is comprised of the
0-, 1-, 2- and 3-particle sectors. According to
papers~\cite{k1,k2}, the sectors with even and odd particle numbers
do not mix together under the action of $S$
and, moreover, in the even sector the $S$-operator
acts as an identical unity.
The 1- and 3-particle sectors do mix together, but it turns out
that {\em there are two eigenvectors of the $S$-operator in the
one-particle sector\/}, with eigenvalues $1$ and $-1$.
Their explicit form can be extracted out of the end
of p.~94 and the beginning of p.~95 of~\cite{k2}.
Namely, denote as $(x,y,z)^{\rm T}$ a one-particle state describing
the situation when the ``amplitudes'' for a particle to be in the
1st, 2nd and 3rd spaces are $x$, $y$ and $z$. According to~\cite{k2},
and taking into account the fact that we are considering vectors
without the 3-particle component, the two vectors
$(x_{\pm},y_{\pm},z_{\pm})^{\rm T}$ with eigenvalues $\pm 1$
are ``isotropic'', i.e. such that $x_{\pm}^2-y_{\pm}^2+z_{\pm}^2=0$,
and in fact it follows from~\cite{k2} that we can restore the $S$-operator
out of {\em any\/} two given vectors of this kind.

Any linear combination $(x,y,z)^{\rm T}$ of the vectors
$(x_{\pm},y_{\pm},z_{\pm})^{\rm T}$ satisf\/ies, of course, some linear
restriction of the form
\be
y=ax+bz,
\label{3}
\ee
where $a$ and $b$ are easily expressed via $x_{\pm}$, $y_{\pm}$ and
$z_{\pm}$.

\subsection{One-particle states}
\label{sec-onep}

Consider now the whole kagome lattice, which will be assumed inf\/inite
in all directions, unless the contrary is stated explicitly.
For it, the one-particle space
is the direct sum of one-particle spaces over all its vertices
multiplied by vacuums in other places. To indicate a one-particle
vector $\varphi$ means to attach a number --- amplitude $\varphi_A$ --- to
each vertex $A$ of the kagome lattice.
In order to ensure that the vector never comes out
of the one-particle space when we apply to it the transfer matrix
any number of times,
we must take it to be an {\em eigenvector\/} of the
transfer matrix. We do not f\/ix here the exact def\/inition of transfer
matrix (see Subsection~\ref{V-sec-disp} for three variants of it),
but note that we must properly take into account
the fact that the transfer matrix
turns inside out half of the triangles of the kagome lattice,
thus moving the lines.

So, let us write down the conditions for a vector in the one-particle space
to be an eigenvector. Consider the following picture (Fig.~\ref{fig1})
representing a fragment of the kagome lattice.
Here the triangle $DCE$ is going to be turned inside out, while
the triangle $BCA$ has been obtained by turning inside out a triangle
on the previous step. So, the two conditions arise:
\be
\varphi_C=a\varphi_D+b\varphi_E
\label{4}
\ee
and
\be
\varphi_C=a\varphi_B+b\varphi_A
\label{5}
\ee
(compare~(\ref{3})).
\begin{figure}[!ht]
\begin{center}
\unitlength=1.00mm
\linethickness{0.4pt}
\begin{picture}(50.00,50.00)
\put(10.00,0.00){\line(3,5){30.00}}
\put(40.00,0.00){\line(-3,5){30.00}}
\put(0.00,45.00){\line(1,0){50.00}}
\put(50.00,5.00){\line(-1,0){50.00}}
\put(8.00,47.00){\makebox(0,0)[rb]{$A$}}
\put(42.00,47.00){\makebox(0,0)[lb]{$B$}}
\put(42.00,3.00){\makebox(0,0)[lt]{$E$}}
\put(8.00,3.00){\makebox(0,0)[rt]{$D$}}
\put(29.00,25.00){\makebox(0,0)[lc]{$C$}}
\put(25.00,22.00){\vector(0,-1){5.00}}
\put(13.00,5.00){\circle*{1.00}}
\put(13.00,45.00){\circle*{1.00}}
\put(37.00,45.00){\circle*{1.00}}
\put(37.00,5.00){\circle*{1.00}}
\put(25.00,25.00){\circle*{1.00}}
\put(34.00,7.00){\vector(-4,3){5.00}}
\put(16.00,7.00){\vector(4,3){5.00}}
\end{picture}
\end{center}

\vspace{-3mm}

\caption{}
\label{fig1}
\end{figure}

When the triangle $DCE$ is turned inside out, it yields
a triangle $D'C'E'$ (Fig.~\ref{fig2}),
and the new ``f\/ield'' variables
are expressed through the old ones as
\be
\pmatrix{\varphi_{D'}\cr \varphi_{E'}}=
\pmatrix{\alpha & \beta \cr \gamma & \delta}
\pmatrix{\varphi_D\cr \varphi_E},
\label{6}
\ee
where it can be derived from the above that
\be
\alpha=-\delta,\qquad \alpha^2+\beta\gamma=1.
\label{6.5}
\ee
\begin{figure}[!ht]
\begin{center}
\unitlength=1.00mm
\linethickness{0.4pt}
\begin{picture}(50.00,30.00)
\put(0.00,25.00){\line(1,0){50.00}}
\put(8.00,27.00){\makebox(0,0)[rb]{$E'$}}
\put(42.00,27.00){\makebox(0,0)[lb]{$D'$}}
\put(29.00,5.00){\makebox(0,0)[lc]{$C'$}}
\put(25.00,14.00){\vector(0,-1){5.00}}
\put(22.00,0.00){\line(3,5){18.00}}
\put(28.00,0.00){\line(-3,5){18.00}}
\put(25.00,5.00){\circle*{1.00}}
\put(37.00,25.00){\circle*{1.00}}
\put(13.00,25.00){\circle*{1.00}}
\put(22.00,19.00){\vector(-4,3){5.00}}
\put(28.00,19.00){\vector(4,3){5.00}}
\end{picture}
\end{center}

\vspace{-3mm}

\caption{}
\label{fig2}
\end{figure}

The points $E'$ and $D'$ of the ``new'' lattice are analogs of
the points $A$ and $B$ correspondingly belonging to the ``old'' lattice.
Thus, in order to obtain on the new lattice a vector proportional
to the vector on the old lattice, we must require that
\be
{\varphi_{E'}\over \varphi_{D'}}={\varphi_A\over \varphi_B}.
\label{7}
\ee
If this condition holds, one can extend
both the old vector $\varphi$ and the new ``primed'' vector periodically
onto the whole lattice and in such way that the new one will
be proportional to the (shifted) old one.

The condition~(\ref{7}) together with (\ref{4}), (\ref{5}), (\ref{6}) is enough
to obtain $\varphi_A$ and $\varphi_B$ (as well as $\varphi_C$,
$\varphi_{D'}$ and $\varphi_{E'}$) out of given $\varphi_D$ and
$\varphi_E$. Thus, only one essential free parameter, e.g. 
$\varphi_D / \varphi_E$, remains
for our construction of one-particle eigenvectors.

\subsection{Two-particle states}
\label{sec-twop}

How can the superposition of two one-particle states of
Subsection~\ref{sec-onep} look like? The experience of studying
the $2+1$-dimensional {\em classical\/} integrable models hints that
probably the ``scattering'' of two particles on one another
must be trivial, i.e. it makes sense to assume
for the ``amplitude of the event that two particles are in two dif\/ferent
points $F$ and $G$ of the lattice'' the form
\be
\Phi_{FG}=\varphi_F \psi_G + \varphi_G \psi_F,
\label{8}
\ee
where $\varphi_{\ldots}$ and $\psi_{\ldots}$ are one-particle amplitudes
like those constructed in Subsection~\ref{sec-onep}.

To see how $\Phi_{FG}$ transforms under the action of transfer matrix,
let us decompose $\varphi_F$ and $\psi_F$, considered
as functions of $F$, in sums
over triangles of the type $DCE$ in Fig.~\ref{fig1}, i.e.
represent $\varphi_F$ and $\psi_F$ as sums of summands each of which
equals zero if $F$ lies beyond the corresponding triangle.
In this way, $\Phi_{FG}$ naturally decomposes in a sum over (non-ordered)
pairs of such triangles, including pairs of two coinciding triangles.
We want that $\Phi_{FG}$ be transformed by the transfer matrix in an
expression of the same form~(\ref{8}), with $\varphi_{\ldots}$ and
$\psi_{\ldots}$ changed to their images with respect to this action.

It is easy to see that this holds automatically if $F$ and $G$ belong
to {\em different\/} triangles. So, it remains to consider the case
where $F$ and $G$  belong to the same triangle, say triangle~$DCE$
in Fig.~\ref{fig1}. When this triangle is transformed by the transfer
matrix in the triangle~$D'C'E'$ of Fig.~\ref{fig2}, the one-particle
amplitudes are transformed according to~(\ref{6}):
\be
\pmatrix{\varphi_{D'}\cr \varphi_{E'}}=
\pmatrix{\alpha & \beta \cr \gamma & \delta}
\pmatrix{\varphi_D\cr \varphi_E},
\qquad
\pmatrix{\psi_{D'}\cr \psi_{E'}}=
\pmatrix{\alpha & \beta \cr \gamma & \delta}
\pmatrix{\psi_D\cr \psi_E},
\label{9}
\ee
where one must add the conditions of type~(\ref{4}):
\be
\ba{ll}
\varphi_C=a\varphi_D+b\varphi_E, \qquad &
\varphi_{C'}=a\varphi_{D'}+b\varphi_{E'}, \vspace{2mm}\\
\psi_C=a\psi_D+b\psi_E, & \psi_{C'}=a\psi_{D'}+b\psi_{E'}. 
\ea \label{10}
\ee
On the other hand, the $S$-matrix (\ref{matrica S 1})--(\ref{matrica S 4})
acts trivially,
i.e. as a unity matrix, in the 2-particle sector. Thus, it must be
\be
\Phi_{C'D'}=\Phi_{CD},\qquad
\Phi_{C'E'}=\Phi_{CE},\qquad
\Phi_{D'E'}=\Phi_{DE}.
\label{11}
\ee
Together, the formulae (\ref{10}), (\ref{11}) and (\ref{8})
lead to the following conditions on the one-particle amplitudes
$\varphi_{\ldots}$ and $\psi_{\ldots}$:
\be
\varphi_{E'}\psi_{D'}+\varphi_{D'}\psi_{E'}=
\varphi_E\psi_D+\varphi_D\psi_E,\label{12}
\ee
\be
\varphi_{D'}\psi_{D'}=\varphi_D\psi_D,
\ee
\be
\varphi_{E'}\psi_{E'}=\varphi_E\psi_E. \label{14}
\ee

The three conditions (\ref{12})--(\ref{14}) together with~(\ref{9})
and~(\ref{6.5}) give, remarkably, just {\em one\/} condition
\be
-\gamma\varphi_D\psi_D + \alpha(\varphi_D\psi_E+\varphi_E\psi_D) +
\beta\varphi_E\psi_E = 0
\label{15}
\ee
on $\varphi_{\ldots}$ and $\psi_{\ldots}$. Recall that, according
to Subsection~\ref{sec-onep}, each of the vectors $\varphi$ and $\psi$~is 
parametrized by one parameter (besides a trivial scalar factor).
Together they are pa\-ra\-metrized by two parameters, but
condition~(\ref{15}) subtracts
one parameter. Thus, the two-particle eigenstates constructed
in this subsection depend on one signif\/icant parameter.

\subsection{Dispersion relations}
\label{V-sec-disp}

The constructed eigenvectors of transfer matrix~$T$ are of course also
eigenvectors for translation operators through periods of kagome
lattice. Let us consider here relations between
the corresponding eigenvalues for the one-particle eigenstate.

It will be convenient to deform slightly the kagome lattice and imagine
it as consisting of horizontal, oblique and vertical lines.
Consider once again some triangle $ABC$ of the kagome lattice,
and its image $A'B'C'$ under the action of $S$-matrix-hedgehog,
as in Fig.~\ref{figV-3-1}.
\begin{figure}[!ht]
\begin{center}
\unitlength=1.00mm
\special{em:linewidth 0.4pt}
\linethickness{0.4pt}
\begin{picture}(65.00,32.50)
\put(3.00,12.00){\line(1,0){20.00}}
\put(23.00,12.00){\line(0,1){20.00}}
\put(23.00,32.00){\line(-1,-1){20.00}}
\put(30.00,17.00){\vector(1,0){6.00}}
\put(43.00,2.00){\line(0,1){20.00}}
\put(43.00,22.00){\line(1,0){20.00}}
\put(63.00,22.00){\line(-1,-1){20.00}}
\put(3.00,12.00){\circle*{1.00}}
\put(23.00,12.00){\circle*{1.00}}
\put(23.00,32.00){\circle*{1.00}}
\put(43.00,22.00){\circle*{1.00}}
\put(63.00,22.00){\circle*{1.00}}
\put(43.00,2.00){\circle*{1.00}}
\put(1.00,12.00){\makebox(0,0)[rc]{$A$}}
\put(25.00,32.00){\makebox(0,0)[lc]{$C$}}
\put(41.00,2.00){\makebox(0,0)[rc]{$C'$}}
\put(65.00,22.00){\makebox(0,0)[lc]{$A'$}}
\put(41.00,22.00){\makebox(0,0)[rc]{$B'$}}
\put(25.00,12.00){\makebox(0,0)[lc]{$B$}}
\end{picture}
\end{center}

\vspace{-6mm}

\caption{}
\label{figV-3-1}
\end{figure}
Let us write out some relations of type (\ref{6}), namely
\be
\pmatrix{\varphi_{A'} \cr \varphi_{B'}}=
\pmatrix{a&b\cr c&d} \pmatrix{\varphi_A \cr \varphi_B},
\label{V-3-1}
\ee
\be
\pmatrix{\varphi_{B'} \cr \varphi_{C'}}=
\pmatrix{\tilde a&\tilde b\cr \tilde c&\tilde d}
\pmatrix{\varphi_B \cr \varphi_C},
\label{V-3-2}
\ee
where $\varphi_{\ldots}$ is any one-particle vector, and the numbers
$a, \ldots ,\tilde d$ (now playing the role of Greek letters
in (\ref{6})) satisfy conditions of type~(\ref{6.5}), i.e.\
\[
\ba{l}
a=-d, \qquad ad-bc=-1, \vspace{1mm}\\
\tilde a=-\tilde d, \qquad \tilde a\tilde d-\tilde b\tilde c=-1.
\ea
\]
 From (\ref{V-3-1}) follows
\be
{\varphi_B\over \varphi_{B'}}=
{-a(\varphi_A/\varphi_{A'})+1\over (\varphi_A/\varphi_{A'})-a},
\label{V-3-3}
\ee
and from (\ref{V-3-2}) follows
\[
{\varphi_C\over \varphi_{C'}}=
{-\tilde a(\varphi_B/\varphi_{B'})+1
\over (\varphi_B/\varphi_{B'})-\tilde a}.
\]
Surely, the numbers $a$ and $\tilde a$ depend on an
$S$-operator-hedgehog. On the other hand, this latter is parameterized
by exactly two parameters. So, in this Subsection we will take
$a$ and $\tilde a$ as those parameters.

We can take for eigenvalue of the hedgehog transfer matrix~$T$
either $\varphi_{A'}/\varphi_A$, or $\varphi_{B'}/\varphi_B$,
or $\varphi_{C'}/\varphi_C$.
These variants correspond, strictly speaking, to dif\/ferent def\/initions
of~$T$, but each of them is consistent with the requirement that
the degrees of~$T$ must be represented graphically as ``oblique
layers'' of cubic lattice (the dif\/ference being that, with the three
dif\/ferent def\/initions, the action of transfer matrix~$T$ corresponds
to the shifts through cubic lattice periods along three dif\/ferent axes).
Our goal is to express the eigenvalues of translation operators
acting {\em within\/} the kagome lattice for a given one-particle state
through, say, $\varphi_{A'}/\varphi_A$.

\begin{figure}[!ht]
\begin{center}
\unitlength=1mm
\linethickness{0.4pt}
\begin{picture}(50.00,50.00)
\put(25.00,0.00){\line(0,1){50.00}}
\put(0.00,25.00){\line(1,0){50.00}}
\put(0.00,20.00){\line(1,1){30.00}}
\put(20.00,0.00){\line(1,1){30.00}}
\put(4.00,27.00){\makebox(0,0)[rb]{$A$}}
\put(23.00,46.00){\makebox(0,0)[rb]{$C$}}
\put(46.00,23.00){\makebox(0,0)[lt]{$D$}}
\put(26.00,3.00){\makebox(0,0)[lt]{$E$}}
\put(27.00,27.00){\makebox(0,0)[lb]{$B$}}
\put(5.00,25.00){\circle*{1.00}}
\put(25.00,5.00){\circle*{1.00}}
\put(25.00,25.00){\circle*{1.00}}
\put(25.00,45.00){\circle*{1.00}}
\put(45.00,25.00){\circle*{1.00}}
\end{picture}
\end{center}

\vspace{-5mm}

\caption{}
\label{figV-3-2}
\end{figure}

If we speak about translation through one lattice
period {\em to the right\/}
in the sense of Figs.~\ref{figV-3-1} and~\ref{figV-3-2},
then this eigenvalue is $\varphi_D/\varphi_A$. It is clear that
\[
{\varphi_D\over \varphi_B}={\varphi_{A'}\over \varphi_{B'}}
\]
--- the ratios of values $\varphi_{\ldots}$ in the triangle $DBE$ are
the same as in $A'B'C'$. Thus,
\be
{\varphi_D\over \varphi_A}={\varphi_{A'}\over \varphi_{B'}}
{\varphi_B\over \varphi_A}={\varphi_{A'}\over \varphi_A} \cdot
{-a(\varphi_A/\varphi_{A'})+1 \over (\varphi_A/\varphi_{A'})-a}
\label{V-3-6}
\ee
(we have used (\ref{V-3-3}). A similar relation can be written out
for the translation through one lattice period in {\em upward\/}
direction in the sense of Figs.~\ref{figV-3-1} and~\ref{figV-3-2},
namely
\be
{\varphi_C\over \varphi_E}=
{\varphi_{B'}\over \varphi_B} \cdot
{-\tilde a(\varphi_B/\varphi_{B'})+1 \over
(\varphi_B/\varphi_{B'})-\tilde a},
\label{V-3-7}
\ee
where one has to substitute the expression (\ref{V-3-3})
for $\varphi_B/\varphi_{B'}$.

\setcounter{equation}{0}

\section{Algebraization: a creation operator}
\label{II}

In this section we will provide
a more ``algebraic'' construction of one-particle states,
resembling the $1+1$-dimensional algebraic Bethe ansatz.

Let us give some def\/initions and remarks. We will depict
the operators graphically in such a way that each operator
will have some number of ``incoming edges'' and the same number
of ``outgoing edges'' (or ``links''). To each edge corresponds
its own copy of a two-dimensional complex linear space,
and to several edges of the same (incoming or outgoing) kind
together corresponds the tensor product of their spaces.
Each of the mentioned two-dimensional spaces has a basis
of a {\em 0-particle\/} and {\em 1-particle\/} vectors.

For any, maybe inf\/inite, collection of edges, we will def\/ine
this {\em collection's 0-particle vector}, or {\em vacuum}, as
the tensor product of 0-particle vectors throughout the collection
(in this paper, the meaning of inf\/inite tensor products will
be always clear). Further, we will identify
a 1-particle vector in an edge with its tensor product with
the 0-particle vectors in all the collection's other edges and
def\/ine a {\em collection's 1-particle vector\/} as a formal sum
over all its edges of the corresponding 1-particle vectors,
with any complex coef\/f\/icients. Then, we can def\/ine in an obvious
way the 2-particle, 3-particle etc. states.

So, according to the above, we assume in this Section that
an operator acts from
the tensor product of ``incoming'' spaces to the tensor product
of ``outgoing'', i.e. dif\/ferent, spaces.
Still, sometimes we will assume that all the
edges along one straight line represent {\em the same\/}
two-dimensional space. This is convenient e.g.\ when we write
out the tetrahedron equation, as in formula~(\ref{tetr}),
and this will never lead to confusion.

We are going to present a one-parameter family of
``creation operators''. When applied to the ``vacuum'', these
operators produce one-particle states --- plane waves, described
in Section~\ref{I}. As we will see, the very construction of these
operators presupposes that they act on vectors which don't dif\/fer
much from the ``vacuum''. We will not try to make this statement
more exact here. Instead, in this Section it will be enough for us
that the domain of def\/inition of those operators contains the
one-dimensional space generated by vacuum.

\subsection{Description of transfer matrices from which the creation
 operators are constructed}
\label{subsec-1}

Creation operators will be transfer matrices on a kagome lattice
with some special boundary conditions. Graphically, such a transfer
matrix is depicted in Fig.~\ref{figII-1}.
As we are considering the eigenstates of transfer matrix~$T$
made up of ``hedgehogs'', as in Section~\ref{I}, the
kagome transfer matrix must be such that
it should be possible to bring the hedgehogs through it using
the tetrahedron equation.

\begin{figure}[!ht]
\begin{center}
\unitlength=1.00mm
\linethickness{0.4pt}
\begin{picture}(120.00,50.00)
\put(25.00,0.00){\line(1,1){50.00}}
\put(45.00,0.00){\line(1,1){50.00}}
\put(65.00,0.00){\line(1,1){50.00}}
\put(35.00,15.00){\line(1,0){50.00}}
\put(55.00,35.00){\line(1,0){50.00}}
\put(20.00,0.00){\line(2,1){100.00}}
\put(40.00,0.00){\line(2,1){60.00}}
\put(60.00,0.00){\line(2,1){20.00}}
\put(40.00,20.00){\line(2,1){60.00}}
\put(60.00,40.00){\line(2,1){20.00}}
\put(30.00,2.00){\line(0,1){6.00}}
\put(50.00,2.00){\line(0,1){6.00}}
\put(70.00,2.00){\line(0,1){6.00}}
\put(40.00,12.00){\line(0,1){6.00}}
\put(50.00,12.00){\line(0,1){6.00}}
\put(60.00,12.00){\line(0,1){6.00}}
\put(70.00,12.00){\line(0,1){6.00}}
\put(80.00,12.00){\line(0,1){6.00}}
\put(50.00,22.00){\line(0,1){6.00}}
\put(70.00,22.00){\line(0,1){6.00}}
\put(90.00,22.00){\line(0,1){6.00}}
\put(60.00,32.00){\line(0,1){6.00}}
\put(70.00,32.00){\line(0,1){6.00}}
\put(80.00,32.00){\line(0,1){6.00}}
\put(90.00,32.00){\line(0,1){6.00}}
\put(100.00,32.00){\line(0,1){6.00}}
\put(70.00,42.00){\line(0,1){6.00}}
\put(90.00,42.00){\line(0,1){6.00}}
\put(110.00,42.00){\line(0,1){6.00}}
\end{picture}

\vspace{-5mm}

\end{center}
\caption{}
\label{figII-1}
\end{figure}

\begin{figure}[!ht]
\begin{center}
\unitlength=1.00mm
\linethickness{0.4pt}
\begin{picture}(120.00,120.00)
\thicklines
\put(0.00,10.00){\line(1,0){89.00}}
\put(91.00,10.00){\line(1,0){18.00}}
\put(111.00,10.00){\line(1,0){9.00}}
\put(0.00,50.00){\line(1,0){49.00}}
\put(51.00,50.00){\line(1,0){18.00}}
\put(69.00,50.00){\line(0,0){0.00}}
\put(71.00,50.00){\line(1,0){49.00}}
\put(0.00,90.00){\line(1,0){9.00}}
\put(11.00,90.00){\line(1,0){18.00}}
\put(31.00,90.00){\line(1,0){89.00}}
\put(30.00,0.00){\line(0,1){69.00}}
\put(30.00,71.00){\line(0,1){18.00}}
\put(30.00,91.00){\line(0,1){29.00}}
\put(70.00,0.00){\line(0,1){29.00}}
\put(70.00,31.00){\line(0,1){18.00}}
\put(70.00,51.00){\line(0,1){69.00}}
\put(110.00,0.00){\line(0,1){9.00}}
\put(110.00,11.00){\line(0,1){109.00}}
\put(0.00,80.00){\line(1,1){9.00}}
\put(11.00,91.00){\line(1,1){29.00}}
\put(0.00,40.00){\line(1,1){29.00}}
\put(31.00,71.00){\line(1,1){49.00}}
\put(0.00,0.00){\line(1,1){49.00}}
\put(51.00,51.00){\line(1,1){69.00}}
\put(40.00,0.00){\line(1,1){29.00}}
\put(71.00,31.00){\line(1,1){49.00}}
\put(80.00,0.00){\line(1,1){9.00}}
\put(91.00,11.00){\line(1,1){29.00}}
\put(10.00,90.00){\circle{2.00}}
\put(30.00,90.00){\circle{2.00}}
\put(30.00,70.00){\circle{2.00}}
\put(50.00,50.00){\circle{2.00}}
\put(70.00,50.00){\circle{2.00}}
\put(70.00,30.00){\circle{2.00}}
\put(90.00,10.00){\circle{2.00}}
\put(110.00,10.00){\circle{2.00}}
\put(10.00,90.00){\makebox(0,0)[cc]{.}}
\put(30.00,90.00){\makebox(0,0)[cc]{.}}
\put(30.00,70.00){\makebox(0,0)[cc]{.}}
\put(50.00,50.00){\makebox(0,0)[cc]{.}}
\put(70.00,50.00){\makebox(0,0)[cc]{.}}
\put(70.00,30.00){\makebox(0,0)[cc]{.}}
\put(90.00,10.00){\makebox(0,0)[cc]{.}}
\put(110.00,10.00){\makebox(0,0)[cc]{.}}
\put(-2.00,93.00){\makebox(0,0)[lb]{$B$}}
\put(22.00,108.00){\makebox(0,0)[lb]{$D$}}
\put(113.00,18.00){\makebox(0,0)[lb]{$C$}}
\put(88.00,3.00){\makebox(0,0)[lb]{$A$}}
\put(30.00,96.00){\vector(0,1){0.00}}
\put(5.00,90.00){\vector(1,0){0.00}}
\put(8.00,88.00){\vector(1,1){0.00}}
\put(28.00,68.00){\vector(1,1){0.00}}
\put(30.00,65.00){\vector(0,1){0.00}}
\put(45.00,50.00){\vector(1,0){0.00}}
\put(48.00,48.00){\vector(1,1){0.00}}
\put(68.00,28.00){\vector(1,1){0.00}}
\put(70.00,25.00){\vector(0,1){0.00}}
\put(85.00,10.00){\vector(1,0){0.00}}
\put(88.00,8.00){\vector(1,1){0.00}}
\put(36.00,90.00){\vector(1,0){0.00}}
\put(70.00,56.00){\vector(0,1){0.00}}
\put(76.00,50.00){\vector(1,0){0.00}}
\put(110.00,16.00){\vector(0,1){0.00}}
\put(103.00,10.00){\vector(1,0){0.00}}
\put(100.00,20.00){\vector(1,1){0.00}}
\put(80.00,40.00){\vector(1,1){0.00}}
\put(70.00,43.00){\vector(0,1){0.00}}
\put(63.00,50.00){\vector(1,0){0.00}}
\put(60.00,60.00){\vector(1,1){0.00}}
\put(40.00,80.00){\vector(1,1){0.00}}
\put(30.00,83.00){\vector(0,1){0.00}}
\put(23.00,90.00){\vector(1,0){0.00}}
\put(20.00,100.00){\vector(1,1){0.00}}
\thinlines
\multiput(87.00,3.00)(-5.00,5.00){18}{\line(-1,1){4.00}}
\multiput(112.00,18.00)(-5.00,5.00){18}{\line(-1,1){4.00}}
\end{picture}
\end{center}

\vspace{-5mm}

\caption{}
\label{figII-2}
\end{figure}

In the tetrahedron equation~(\ref{tetr})
a number 0, 1, 2, or 3 is attached to a {\em plane}, that is,
to a face of the tetrahedron. An operator $S_{ij,ik,jk}$ acts
in the tensor product of three linear spaces corresponding
to the {\em lines} --- intersections of those planes.

Let us assume that parameters $\varphi_1,\varphi_2,\varphi_3$ belong
to the ``hedgehogs'' $S_{12,13,23}$ and are given.
Then we will build the kagome transfer matrix out of matrices
$S_{01,02,12}$, $S_{01,03,13}$ and $S_{02,03,23}$ in such way
that with the help of (\ref{tetr}) one could pass a hedgehog
through the kagome lattice. Here the number 0 (and the corresponding
parameter~$\varphi_0$) is attached to the plane
of the kagome lattice.

\subsection{Boundary conditions for creation operators}

It will take some ef\/fort to describe the boundary conditions that we
are going to impose on kagome transfer matrices of
Subsection~\ref{subsec-1} to obtain out of them creation operators.
The problem is that we are considering a kagome lattice {\em infinite\/}
in all plane directions. So, f\/irst, let us draw
in Fig.~\ref{figII-2}
the lattice viewed from above.
Then let us draw a dashed line $AB$ and cut of\/f for a while
the part of the lattice lying to the left of that line
(it will be explained in Subsection~\ref{subsec-calc}
that really there is much arbitrariness in choosing the line $AB$,
but let it be for now as in Fig.~\ref{figII-2}). For the rest
of transfer matrix, let us def\/ine the boundary condition along $AB$
as follows. Consider all the lattice edges intersecting $AB$.
They are {\em incoming edges\/} for the remaining part of transfer
matrix. To def\/ine the boundary conditions, we must indicate some vector
$\Sigma_{AB}$ in the tensor product of corresponding spaces. Let us assume
that $\Sigma_{AB}$ is a {\em 1-particle vector\/} as def\/ined
in the beginning of this Section,
whose exact form is to be determined.

Consider the band --- the part of transfer matrix lying between
the lines $AB$ and $CD$. This band represents an operator acting from
the space corresponding to its incoming edges into the space
corresponding to its outgoing edges, where the incoming edges
are those intersecting $AB$ and also those pointing from behind the kagome
lattice plane into the vertices situated within the band,
while the outgoing edges are those intersecting $CD$
and those pointing from the vertices situated within the band
at the reader. The latter vertices
are marked $\alh$ in Fig.~\ref{figII-2}.

Let us require that our band operator --- let us call it $\cal B$ --- transform
the tensor product
$\Sigma_{AB} \otimes \Omega_{\alh}$, where $\Omega_{\alh}$
is the vacuum for the set of edges pointing into the ``$\alh$'' vertices,
into the following sum:
\be
{\cal B}\, \Sigma_{AB} \otimes \Omega_{\alh} =
\kappa\,\Sigma_{CD} \otimes \Omega'_{\alh} + \Omega_{CD}\otimes
\Psi_{\alh},
\label{calB}
\ee
where $\kappa$ is a number; $\Sigma_{CD}$ is the vector similar to
$\Sigma_{AB}$, but corresponding to the edges situated one lattice
period to the right, i.e. intersecting $CD$; $\Omega'_{\alh}$ is
the vacuum for edges pointing from the ``$\alh$'' vertices to the reader
(who can thus identify $\Omega'_{\alh}$ with $\Omega_{\alh}$ if desired);
$\Omega_{CD}$ is the vacuum for the set of vectors intersecting $CD$;
$\Psi_{\alh}$ is some vector lying in the same tensor product of
spaces as $\Omega'_{\alh}$.
It is remarkable that, for any $\varphi_0$, relation (\ref{calB})
can be satisf\/ied
with a proper choice of~$\Sigma_{AB}$. The vector $\Psi_{\alh}$ will then
be a 1-particle vector. Some details of calculations concerning
relation~(\ref{calB}) are explained in Subsection~\ref{subsec-calc},
while here we are going to {\em use\/} this relation.

The next band, lying to the right of $CD$, has vector
$\kappa\,\Sigma_{CD}$
as its incoming vector. Thus, the whole situation is repeated up to
the factor~$\kappa$. On the other hand, we could have cut the lattice,
instead of the line $AB$, along some other line lying, e.g.,
$n$ lattice periods to the left. In that case, we should have taken
for the incoming vector the vector $\Sigma_{AB}$ shifted by $n$
periods to the left and multiplied by~$\kappa^{-n}$.
Letting $n\to\infty$, we get a $\Psi_{\alh}$-like vector
in {\em every\/} band of the sort depicted in Fig.~\ref{figII-2}.
Summing up
all those vectors, we get a 1-particle state of the same kind as in
Section~\ref{I}. Those states are now parameterized by
the parameter~$\varphi_0$.

In fact, one more boundary condition must be imposed at the
``right inf\/inity'' of the lattice. This is explained in the end of
Subsection~\ref{subsec-calc}.

\subsection{Some technical details}
\label{subsec-calc}

Instead of the straight line $AB$ in Fig.~\ref{figII-2}, we could use
any (connected) curve $l$ {\em intersecting each straight line of the kagome
lattice exactly one time\/} in such way that a boundary condition
is given in the tensor product corresponding to the edges that
intersect~$l$. Assuming that a 1-particle vector is given as the boundary
condition, we will require that after any deformation of~$l$ such that
it passes through {\em one\/} of the lattice vertices, the boundary
condition remain to be 1-particle.

In other words, the mentioned vertex is added to or withdrawn from
the considered part of the lattice. Let that vertex be, e.g., such as
in Fig.~\ref{figII-3}.
An incoming 1-particle vector for it is described
by two amplitudes $a$ and $b$, with $a$ corresponding to the edge~$01$
(see Subsection~\ref{subsec-1}) and $b$ --- to the edge~$02$.
It is required that the result of transforming this incoming vector
by the matrix~$S$ (\ref{matrica S 1})--(\ref{matrica S 4})
contain no three-particle part. This leads at once to the condition
\be
{a\over b}=-\,\root\of{\tanh(\varphi_0-\varphi_1)}
 \,\,\,\root\of{\tanh(\varphi_0-\varphi_2)}.
\ee

\begin{figure}[!ht]
\vspace{-5mm}

\begin{center}
\unitlength=1mm
\linethickness{0.4pt}
\begin{picture}(40.00,40.00)
\put(0.00,0.00){\line(1,1){19.00}}
\put(21.00,21.00){\line(1,1){19.00}}
\put(0.00,20.00){\line(1,0){19.00}}
\put(21.00,20.00){\line(1,0){19.00}}
\put(0.00,0.00){\vector(1,1){12.00}}
\put(0.00,20.00){\vector(1,0){12.00}}
\put(23.00,20.00){\vector(1,0){7.00}}
\put(23.00,23.00){\vector(1,1){7.00}}
\put(20.00,20.00){\circle{2.00}}
\put(20.00,20.00){\makebox(0,0)[cc]{.}}
\put(3.00,21.00){\makebox(0,0)[cb]{$a$}}
\put(3.00,5.00){\makebox(0,0)[cb]{$b$}}
\put(35.00,22.00){\makebox(0,0)[cb]{$f$}}
\put(35.00,37.00){\makebox(0,0)[cb]{$c$}}
\end{picture}

\vspace{-8mm}

\end{center}
\caption{}
\label{figII-3}
\end{figure}

The other ratios of the amplitudes written out in Fig.~\ref{figII-3}
are simply matrix elements of~$S$:
\be
{c\over a}={f\over b}=\,\root\of{\tanh(\varphi_0-\varphi_1)}
 \,\,\,\root\of{\coth(\varphi_0-\varphi_2)}.
\ee

Similar relations can be written for the kagome lattice vertices
of two other kinds, that is
$\matrix{
\unitlength=1mm
\linethickness{0.4pt}
\begin{picture}(6.00,7.00)
\put(0.00,3.50){\line(1,0){6.00}}
\put(3.00,0.50){\line(0,1){6.00}}
\end{picture} }$
and
$\matrix{
\unitlength=1mm
\linethickness{0.4pt}
\begin{picture}(6.00,7.00)
\put(3.00,0.50){\line(0,1){6.00}}
\put(0.00,0.50){\line(1,1){6.00}}
\end{picture} }$.
It turns out that all those relations together determine the amplitudes
at {\em all\/} lattice edges from a given one of them without contradiction.
Thus, the amplitudes for the $\alh$-edges, that are incoming for
the hedgehog transfer matrix, are determined correctly.

Those amplitudes give exactly its eigenstate for any f\/ixed~$\varphi_0$.
This can be proved by a rather obvious reasoning:
use the tetrahedron equation and the possibility to express the amplitudes
at dif\/ferent edges through one another. The details are left
for the reader.

There remains, however, another detail that is important:
we must impose one more boundary condition, that is at the
``right inf\/inity'' (see again Fig.~\ref{figII-2}).
In order to obtain the 1-particle vector at the $\alh$-edges,
and no {\em vacuum\/} component, let us take a straight
line $C'D'$ --- like $CD$, but somewhere far to the right --- and
require that the vector in the space corresponding
to edges that intersect $C'D'$ have no 1-particle component.
Then, of course, we let $C'D'$ tend to the right inf\/inity,
so this procedure does not change the 1-particle component
of the $\alh$-vector, but the vacuum component vanishes.

\setcounter{equation}{0}

\section{Tetrahedral Zamolodchikov algebras and string-like states}
\label{III}

In this section we show how to construct some string-like eigenstates
using tetrahedral Zamolodchikov algebras. We will be using
the trigonometrical tetrahedral Zamolodchikov algebras
described in~\cite{k-trig}.
The idea is that sometimes we can control the evolution under the action
of transfer matrix powers for the states arising from a (kagome)
lattice of f\/ive-legged `$R$-operators', with given boundary conditions.

Let us return to Fig.~\ref{figIII-1}.
Suppose we have f\/ixed some boundary conditions in the tensor product
of spaces denoted 1, 2 and~3 (which means, most generally, that we have
taken the trace of a product of each side of~(\ref{tza}) and some
linear operator acting in the mentioned tensor product).
This yields, in the l.h.s. of~(\ref{tza}), some vector in the tensor
product of spaces corresponding to indices $a$, $b$ and $c$, and in the
r.h.s. of~(\ref{tza}) --- the result of $S$-operator action upon
a similar vector. For dif\/ferent boundary conditions, this provides
enough (consistent) relations for $S$-operator to be determined
uniquely.

We will look at this, however, from another point of view, using
boundary conditions for $R$'s as a means to def\/ine vectors in the
space where $S$'s act. Of course, we will
take, instead of just three $R$-operators,
a large lattice made up of them, whose fragment is depicted
in Fig.~\ref{figIII-2},
and apply a layer of $S$-operators --- a hedgehog transfer matrix --- to it.
We will see that sometimes, for simple boundary
conditions, this can be a reasonable way of describing vectors
on which the transfer matrix acts, as well as results
of such action.

\begin{figure}[!ht]
\begin{center}
\unitlength=1.00mm
\linethickness{0.4pt}
\begin{picture}(120.00,50.00)
\put(25.00,0.00){\line(1,1){50.00}}
\put(45.00,0.00){\line(1,1){50.00}}
\put(65.00,0.00){\line(1,1){50.00}}
\put(35.00,15.00){\line(1,0){50.00}}
\put(55.00,35.00){\line(1,0){50.00}}
\put(20.00,0.00){\line(2,1){100.00}}
\put(40.00,0.00){\line(2,1){60.00}}
\put(60.00,0.00){\line(2,1){20.00}}
\put(40.00,20.00){\line(2,1){60.00}}
\put(60.00,40.00){\line(2,1){20.00}}
\put(30.00,5.00){\line(0,1){3.00}}
\put(50.00,5.00){\line(0,1){3.00}}
\put(70.00,5.00){\line(0,1){3.00}}
\put(40.00,15.00){\line(0,1){3.00}}
\put(50.00,15.00){\line(0,1){3.00}}
\put(60.00,15.00){\line(0,1){3.00}}
\put(70.00,15.00){\line(0,1){3.00}}
\put(80.00,15.00){\line(0,1){3.00}}
\put(90.00,25.00){\line(0,1){3.00}}
\put(70.00,25.00){\line(0,1){3.00}}
\put(50.00,25.00){\line(0,1){3.00}}
\put(60.00,35.00){\line(0,1){3.00}}
\put(70.00,35.00){\line(0,1){3.00}}
\put(80.00,35.00){\line(0,1){3.00}}
\put(90.00,35.00){\line(0,1){3.00}}
\put(100.00,35.00){\line(0,1){3.00}}
\put(70.00,45.00){\line(0,1){3.00}}
\put(90.00,45.00){\line(0,1){3.00}}
\put(110.00,45.00){\line(0,1){3.00}}
\end{picture}
\end{center}

\vspace{-5mm}

\caption{}
\label{figIII-2}
\end{figure}

As already stated, the $R$-operators we will be dealing with
in this section are the simplest
possible --- trigonometrical --- ones. In this connection, let us refer
to Theorem~2.3 of the work~\cite{k2} wherefrom it follows that to
an $S$-matrix corresponds a {\em two-parameter\/} family of
triples of $R$-operators. If we restrict ourselves to only trigonometrical
$R$-matrices from~\cite{k-trig}, then there remains a {\em one-parameter\/}
family of those. So, below it is implied that we are constructing
one-parameter families of states for a given transfer matrix.

\subsection{Two kinds of strings in a f\/inite lattice}
\label{sec-finite}

\subsubsection{Eigenstates with eigenvalue~1 yielded by the lattice\\
with a given ``polarization''}
\label{topological}

For a f\/inite lattice on a torus, the ``periodic'' boundary
conditions seem, at f\/irst sight, to be already f\/ixed.
However, trigonometrical $R$-operators of work~\cite{k-trig} conserve
the ``number of particles'', sometimes called also ``polarization''
(because they are very much like the usual 6-vertex model
$L$-operators), and this provides more possibilities.
Namely, the reader can easily verify that the following construction
yields some states that are transformed into themselves by the hedgehog
transfer matrix.

Let us declare some edges of the kagome lattice (Fig.~\ref{figIII-2})
`black' and the others `white' in such a way that the number of black
lines is conserved at each vertex (the incoming edges being situated
below and to the left of the vertex, and the outgoing edges --- above
and to the right). It can be said that such a conf\/iguration of black
edges --- we will call it {\em permitted\/} conf\/iguration --- forms
a cycle belonging to some homology class of the torus.
Let us say that vectors $\pmatrix{0\cr 1}$ correspond to white
edges, and vectors $\pmatrix{1\cr 0}$ --- to black edges.
This selects some matrix element for each $R$-operator,
but as there are really two operators $R^0$ and $R^1$,
this selects a pair of numbers forming a vector in the
two-dimensional space corresponding to a {\em vertical\/} edge
in Fig.~\ref{figIII-2}. The tensor product of such vectors lies
in the space where the transfer matrix acts.
Now let us take a sum of those vectors
over all black edges conf\/igurations belonging
to the same homological class.
It is quite straightforward to see that the transfer matrix transforms
this sum into exactly the same sum.

\subsubsection{How moving strings arise from the lattice
of {\mathversion{bold}$R$}-operators}

A slight modif\/ication of the above construction yields moving
straight strings. Namely, f\/ix arbitrarily some straight
lines of the kagome lattice of Fig.~\ref{figIII-2} and paint black
all the edges belonging to them. Then those lines will move under
the action of transfer matrix, just because they are the intersection lines
of the planes of cubic lattice and the moving plane orthogonal
to vector $(1,1,1)$.

This is quite evident for horizontal and vertical lines, and
the only slightly nontrivial consideration is required for
{\em oblique\/} lines.
To make this clear, let us draw some more pictures.
First, let us interchange l.h.s. and r.h.s. in Fig.~\ref{figIII-1}
and redraw it like the following formula:
\[
S \sum \hbox{conf\/igurations of } \matrix{
\unitlength=1mm
\linethickness{0.4pt}
\begin{picture}(14.00,14.00)
\put(0.00,0.00){\line(1,1){14.00}}
\put(4.00,0.00){\line(0,1){14.00}}
\put(0.00,10.00){\line(1,0){14.00}}
\end{picture}
} = \sum \hbox{conf\/igurations of } \matrix{
\unitlength=1mm
\linethickness{0.4pt}
\begin{picture}(14.00,14.00)
\put(0.00,0.00){\line(1,1){14.00}}
\put(0.00,4.00){\line(1,0){14.00}}
\put(10.00,0.00){\line(0,1){14.00}}
\end{picture}
},
\]
where ``conf\/igurations'' means ``vectors corresponding to
permitted conf\/igurations of three black
edges within a triangle with given `boundary condition' for six
external edges''.
In these terms, Fig.~\ref{figIII-1} itself says only that
\[
S \left( \matrix{
\unitlength=1mm
\linethickness{0.4pt}
\begin{picture}(14.00,14.00)
\put(0.00,0.00){\line(1,1){14.00}}
\end{picture}
} + \matrix{
\unitlength=1mm
\linethickness{0.4pt}
\begin{picture}(14.00,14.00)
\put(0.00,0.00){\line(1,1){4.00}}
\put(4.00,4.00){\line(0,1){6.00}}
\put(4.00,10.00){\line(1,0){6.00}}
\put(10.00,10.00){\line(1,1){4.00}}
\end{picture}
}\right) = \matrix{
\unitlength=1mm
\linethickness{0.4pt}
\begin{picture}(14.00,14.00)
\put(0.00,0.00){\line(1,1){14.00}}
\end{picture}
} + \matrix{
\unitlength=1mm
\linethickness{0.4pt}
\begin{picture}(14.00,14.00)
\put(0.00,0.00){\line(1,1){4.00}}
\put(4.00,4.00){\line(1,0){6.00}}
\put(10.00,4.00){\line(0,1){6.00}}
\put(10.00,10.00){\line(1,1){4.00}}
\end{picture}
}
\]
(here only the black edges are depicted), and not that
\be
S \matrix{
\unitlength=1mm
\linethickness{0.4pt}
\begin{picture}(14.00,14.00)
\put(0.00,0.00){\line(1,1){14.00}}
\end{picture}
} = \matrix{
\unitlength=1mm
\linethickness{0.4pt}
\begin{picture}(14.00,14.00)
\put(0.00,0.00){\line(1,1){14.00}}
\end{picture}
}.
\label{oblique}
\ee
However, (\ref{oblique}) is proved by direct calculation involving
the explicit expressions for matrix elements of $R$-operators.

For horizontal and vertical lines, relations similar to (\ref{oblique})
arise, of course, at once. Straight strings are further discussed
in Subsection~\ref{subsec another}.

\subsection{Another approach to straight strings}
\label{subsec another}

\subsubsection{Straight strings from vacuum vectors}
\label{subsubs straight strings vacuum}

The matrix $S$, according to the work~\cite{k2},
has two families of vacuum vectors.
Here we will restrict ourselves to considering the f\/irst family,
i.e. the vacuum vectors transformed by~$S$ into themselves:
\be
S\bigl(X(\zeta) \otimes Y(\zeta) \otimes
 Z(\zeta)\bigr) =
X(\zeta) \otimes Y(\zeta) \otimes Z(\zeta),
\label{16}
\ee
$\zeta$ being a parameter taking values in an elliptic curve
(compare with formula~(1.12) from~\cite{k2}. Strictly speaking,
in~\cite{k2} we were considering vacuum {\em covectors}, but this
does not make much dif\/ference). What we are going to do
in this Subsubsection can be done with the same success for the second family
as well. Let us note that the particle-like
excitations of Sections \ref{I} and~\ref{II} have been
built using {\em both\/} families of vacuum vectors.

The simplest eigenvectors $\Omega(\zeta)$ of the transfer matrix,
with the eigenvalue 1, are built as follows: f\/ix $\zeta$ and
put in correspondence to each point of type~$A$ (Fig.~\ref{figV-3-1})
of the kagome lattice the vector $X(\zeta)$, to each point of
type~$B$ --- the vector $Y(\zeta)$, and of type~$C$ --- the
vector~$Z(\zeta)$.
Then take the tensor product of all those vectors. The formula~(\ref{16})
shows at once that this is indeed an eigenvector with eigenvalue~1.

A little bit more intricate eigenvectors, for which the eigenvalues
in case of a {\em finite\/} lattice are roots of unity, can be
constructed as follows. It is seen from formulae (2.13)--(2.15)
of paper~\cite{k2}, where enter the values $x$, $y$ and $z$ --- ratios
of two coordinates of vectors $X$, $Y$ and $Z$ respectively, --- that
the triple $X$, $Y$, $Z$ will remain vacuum if one makes one of the following
changes:
\be
(x,y,z)  \to  (x,\,1/y,\,1/z), \label{17} 
\ee
\be
(x,y,z)  \to  (1/x,\,y,\,-1/z), \label{18} 
\ee
\be
(x,y,z)  \to  (-1/x,\,-1/y,\,z). \label{19}
\ee
It can be said that the changes (\ref{17}), (\ref{18}) and (\ref{19})
af\/fect respectively the sides $BC$, $AC$ and $AB$ of the triangle~$ABC$
in Fig.~\ref{figV-3-1}. Obviously, two such changes, if applied successively,
commute with one another.

To construct a vector whose transformation under the action of transfer
matrix is easy to trace, let us act like this: f\/irst, select arbitrarily
some straight lines --- {\em strings\/} --- going along the edges
of the kagome lattice.
Then, take the vector $\Omega(\zeta)$ and change it as follows:
make in each triangle of the type~$DCE$ the transformation(s) of
type (\ref{17})--(\ref{19}) if its corresponding side lies in
a selected line.

The obtained vector --- let us call it $\Theta$ --- goes under the action
of transfer matrix $T$ into a vector of a similar form, but with the
properly shifted lines (the latter, let us recall, result from
the intersection of the cubic lattice faces with a plane perpendicular
to the vector $(1,1,1)$, and move in that plane when the plane itself
moves). An eigenvector of $T$ can be now built in the form
\be
\cdots+\omega^{-1}T^{-1}\Theta+\Theta+\omega T\Theta+\omega^2 T^2 \Theta
+\cdots,
\ee
where in the case of a f\/inite lattice the sum must be f\/inite,
and the number $\omega$ must be a root of unity of a proper degree,
determined by the sizes of the lattice.

\subsubsection{Straight strings as symmetries}
\label{sec-string-sym}

Let us introduce the Pauli matrices
\[
\sigma_1=\pmatrix{0 & 1 \cr 1 & 0}, \qquad
\sigma_2=\pmatrix{0 & -i \cr i & 0}, \qquad
\sigma_3=\pmatrix{1 & 0 \cr 0 & -1},
\]
and the unity matrix
\[
\sigma_0=\pmatrix{1 & 0 \cr 0 & 1}.
\]
Note that the subscripts of these matrices have other meaning than
the subscripts of $S$-matrices in equations like~(\ref{tetr}).

It follows from the explicit form of our $S$-matrix
that it commutes with operators
\be
\sigma_2 \otimes \sigma_2 \otimes \sigma_0, \qquad
\sigma_1 \otimes \sigma_0 \otimes \sigma_2 \qquad \hbox{and} \qquad
\sigma_0 \otimes \sigma_1 \otimes \sigma_1.
\label{II-*}
\ee

If now we select some set of the kagome lattice
horizontal lines (to be exact, of those
depicted in Fig.~\ref{figII-2} as horizontal)
and consider the tensor product of matrices $\sigma_2$ over all
vertices belonging to those lines, then the hedgehog transfer matrix~$T$
will be permutable with that product up to the fact that the lines move
in the lattice plane. This permutability follows
immediately from the fact that $S$ commutes with the f\/irst
of operators~(\ref{II-*}).
Similarly, it is not dif\/f\/icult to formulate the analogous statements
for sets of oblique and vertical lines, using respectively the second
and third of products~(\ref{II-*}).

Using the described symmetries of transfer matrix~$T$, we can,
starting from any state vector~$\Theta_0$ whose evolution under
the action of degrees of~$T$ we can describe in this or that way, build
many new states~$\Theta$
whose evolution we will also be able to describe.

\subsection{`Broken' inf\/inite strings}
\label{sec-lom}

Let us return to the kagome lattice made up of f\/ive-legged
$R$-operators. In this subsection this lattice will be inf\/inite.
As in Subsection~\ref{sec-finite},
we will paint black some edges. Namely, they will form
two horizontal rays, one going to the right and one going to
(or rather coming from) the left, as in Fig.~\ref{figIII-3},
and those rays will be connected by some path going along the lattice
edges and {\em permitted\/} in the sense of Subsubsection~\ref{topological}.
Consider the vector --- the formal inf\/inite tensor product --- corresponding
to such black edge conf\/iguration, and take a sum over {\em all\/}
the permitted paths linking the two rays (in particular, the ends
$A$ and $B$ of the rays can move anywhere to the left and/or to
the right).
\begin{figure}[!ht]

\vspace{-3mm}

\begin{center}
\unitlength=1.00mm
\linethickness{0.4pt}
\begin{picture}(120.00,40.00)
\put(0.00,5.00){\line(1,0){50.00}}
\put(70.00,35.00){\line(1,0){50.00}}
\multiput(52.00,8.00)(2.00,3.00){9}{\makebox(0,0)[cc]{.}}
\put(51.00,4.00){\makebox(0,0)[lt]{$A$}}
\put(69.00,36.00){\makebox(0,0)[rb]{$B$}}
\end{picture}
\end{center}

\vspace{-6mm}

\caption{}
\label{figIII-3}
\end{figure}

Thus, the simplest `broken' strings on the inf\/inite lattice appear. As in
Subsubsection~\ref{subsubs straight strings vacuum},
formal eigenvectors can be built out of translations of such strings.

\setcounter{equation}{0}
\section{String --- particle marriage}
\label{IV}

In Section~\ref{III}, we have been using a kagome lattice made up
of f\/ive-legged `$R$-operators' for string construction.
The idea of this Section
is that probably we will be able to construct more
sophisticated strings using
a kagome lattice made up of six-legged `$S$-operators' instead.
Indeed, it turns out that this is a right way to combine ideas
of previous sections.

Here we will be considering
the kagome lattice inf\/inite in all plane directions.
Let us imagine it again as in Fig.~\ref{figII-1}:
situated in a horizontal plane, having
an $S$-operator in each its vertex and using four of six $S$-operator
legs as lattice edges.
Two other legs of an $S$-operators are vertical.
The $S$-operators form in such way some
{\em auxiliary transfer matrix}.
We will apply some vector --- namely, a ``one-particle''
vector from Section~\ref{I} --- to lower
ends of $S$-operators, while a string will be posed, in a sense,
within the plane, as in Section~\ref{III}.
The result will be new eigenstates for the ``hedgehog'' transfer matrix.

As is known, our transfer matrix (\ref{matrica S 1})--(\ref{matrica S 4})
is made up of $S$-operators that can be described as
the ``static limit'' of a more general construction
due to Sergeev, Mangazeev and Stroganov (SMS), see paper~\cite{mss}
(operators are called there `$R$' instead of `$S$').
On the other hand, the auxiliary transfer matrix will be made of some
{\em other\/} special case of SMS operators described
in Subsections \ref{sec-SMS-spec} and~\ref{sec-gauge} below.

Then in Subsection~\ref{sec-absence} we show that a particle
in absence of a string just vanishes, while their ``marriage''
produces in Subsection~\ref{sec-marriage} new nontrivial strings.

\subsection{Tetrahedron equation: a special case}
\label{sec-SMS-spec}

As is known from the paper~\cite{mss}, the solution of the tetrahedron
equation is parameterized by dihedral angles
$\vartheta_1,\ldots,\vartheta_6$ (of which f\/ive are independent)
between four planes in the three-dimensional euclidean space.
The parameterized equation looks like
\be
\ba{l}
S_{123}(\vartheta_1, \vartheta_2, \vartheta_3)\,
S_{145}(\vartheta_1, \vartheta_4, \vartheta_5)\,
S_{246}(\pi-\vartheta_2, \vartheta_4, \vartheta_6)\,
S_{356}(\vartheta_3, \pi-\vartheta_5, \vartheta_6)
\vspace{2mm}\\
\qquad = S_{356}(\vartheta_3, \pi-\vartheta_5, \vartheta_6)\,
S_{246}(\pi-\vartheta_2, \vartheta_4, \vartheta_6)\,
S_{145}(\vartheta_1, \vartheta_4, \vartheta_5)\,
S_{123}(\vartheta_1, \vartheta_2, \vartheta_3) ,
\label{IV-*}
\ea
\ee
where the subscripts of each $S$ number the spaces where it acts.

Let us imagine that those four planes pass through the center
of a unit sphere. Then an alternative way of parameterizing,
say, the operator $S_{123}$ is by using the edges $a_1$, $a_2$
and $a_3$ of the spherical triangle instead of 
$\vartheta_1$, $\vartheta_2$ and $\vartheta_3$. Each vertex
of the triangle is, naturally, the intersection point of the
sphere and a pair of planes belonging to the operator $S_{123}$.

We will assume that the operator $S_{123}$ in the equation~(\ref{IV-*})
is one of the hedgehogs of the ``hedgehog transfer matrix'',
while the other $S$-operators belong to the auxiliary
kagome transfer matrix. Recall that we are considering the hedgehogs
corresponding to the {\em static limit\/} in terms
of~\cite{mss}. This means that the sides $a_1$, $a_2$ and $a_3$
of the spherical triangle are inf\/initely small.
Let this inf\/initely small spherical triangle be situated
at the north pole of the sphere.

Now let us f\/ix a particular place for the fourth plane (that doesn't
pass through the north pole). Namely, let this plane
{\em intersect the sphere through its
equator}. The resulting restrictions on the angles $\vartheta$
are the following:
\be
\vartheta_1 +\vartheta_2 +\vartheta_3 =\pi, \qquad
\vartheta_4 =\vartheta_5 =\vartheta_6 ={\pi \over 2}.
\label{IV-**}
\ee
We will see that (\ref{IV-**}) yields degenerate
operators $S_{145}$, $S_{246}$ and $S_{356}$,
but this only helps to perform our construction.

\subsection{The gauge for ``doubly rectangular'' {\mathversion{bold}$S$}-opera\-tors}
\label{sec-gauge}

The explicit form for operators entering in equation~(\ref{IV-*})
can be found in section~5 of the work~\cite{mss}.
In the same work, quite a bit is said about dif\/ferent gauges
for those operators. We will work with the following gauge
for our ``doubly rectangular'' --- i.e.\ having
two of three angles~$\vartheta$ equal to $\pi/2$ --- operators:
f\/irst take them as in section~5 of~\cite{mss}, and then change
as described in the following paragraph.

Consider e.g. the operator $S_{145}=S_{145} (\vartheta_1,
\vartheta_4,\vartheta_5)=S_{145} (\vartheta_1, \pi/2, \pi/2)$.
Its legs corresponding to the spaces number 4 and~5 lie on the
edges of the kagome lattice, in the horizontal plane (see
the beginning of this Section).
Let us perform a conjugation in both those spaces
with the matrix $\pmatrix{1&-1\cr 1&1}$, that is, let us change
\[
S_{145}\to {\bf 1}\otimes \pmatrix{1&-1\cr 1&1} \otimes
\pmatrix{1&-1\cr 1&1} \cdot S_{145} \cdot {\bf 1}\otimes
\pmatrix{1&-1\cr 1&1}^{-1} \otimes \pmatrix{1&-1\cr 1&1}^{-1}.
\]

Let us perform the similar transformations for $S_{246}$
in the spaces 4 and~6, and for $S_{356}$ in the spaces 5 and~6.
Note that such gauges are consistent with the tetrahedron
equation~(\ref{IV-*}), and that the gauge of $S_{123}$ does not
change.

Now let us forget about the old gauge of the operators
$S_{145}$, $S_{246}$ and $S_{356}$, and use these notations
for the operators in the {\em new\/} gauge. The explicit form
of each of these operators is like this:
\[
S_{\ldots}= {1-\tan(\vartheta/4)\over \cos(\vartheta/2)}
\pmatrix{A&B\cr C&D},
\]
where
\be
A=\pmatrix{
\cos(\vartheta/2) &0&0&0 \cr
0& \sin (\vartheta/2) & 1 &0 \cr
0& 1 & \sin (\vartheta/2) &0 \cr
0&0&0& \cos (\vartheta/2)
},
\label{IV-A}
\ee
\be
B= {\root\of{\sin\vartheta}\over 2} \pmatrix{
1+i &0&0&0 \cr
0& 1-i &0&0 \cr
0&0& -1+i &0 \cr
0&0&0& -1-i
},
\label{IV-B}
\ee
\be
C= {\root\of{\sin\vartheta}\over 2} \pmatrix{
1-i &0&0&0 \cr
0& 1+i &0&0 \cr
0&0& -1-i &0 \cr
0&0&0& -1+i 
},
\label{IV-C}
\ee
\be
D=\pmatrix{
\sin(\vartheta/2) &0&0& -1 \cr
0& \cos(\vartheta/2) &0&0 \cr
0&0& \cos(\vartheta/2) &0 \cr
-1 &0&0& \sin(\vartheta/2)
},
\label{IV-D}
\ee
where
\[
\ba{l}
\vartheta= \vartheta_1 \qquad \mbox{for} \quad S_{145}, \vspace{1mm}\\
\vartheta= \vartheta_2 \qquad \mbox{for} \quad S_{246}, \vspace{1mm}\\
\vartheta= \vartheta_3 \qquad \mbox{for} \quad S_{356}. 
\ea
\]

Matrix $A$ is the matrix of weights for all conf\/igurations
of horizontal spins of an $S$-operator, if the spins at the vertical
edges are both f\/ixed and equal~0. Similarly, matrix~$B$
corresponds to the lower spin~1 and the upper spin~0;
matrix~$C$ --- to the lower spin~0 and the upper spin~1;
and matrix~$D$ --- to the lower spin~1 and the upper spin~1.

We see that all of these matrices except~$D$ {\em conserve the
total spin}, or, in the other terminology, the ``number of particles'',
within the horizontal plane.

\subsection{Disappearance of particle in absence of a string}
\label{sec-absence}

Now let us apply the auxiliary kagome transfer matrix made up
of operators $S_{145}$, $S_{246}$ and $S_{356}$ to the
one-particle vector from Sections \ref{I} and~\ref{II}.
We will choose the following boundary conditions at the horizontal
inf\/inity: all spins at the ``far enough'' horizontal edges
are zero.

Let us show that in this case the spins at {\em all\/} horizontal edges
are zero. Recall that the one-particle state is a linear combination
of tensor products of spins, of which {\em one\/} equals unity,
and the others are zero. This means that matrix~$D$ --- the only one that can
change the total
spin --- cannot appear twice. In other words, non-zero spin cannot
be created somewhere and then annihilated somewhere else,
thus all the horizontal edges possess zero spins.

This conclusion implies that only {\em upper left\/} entries
of the matrices $A,\ldots,D$ are playing r\^ole. For a given
$S$-operator, four such elements form a matrix
\be
{1-\tan(\vartheta/4)\over \cos(\vartheta/2)} \pmatrix{
\cos(\vartheta/2) & {\textstyle 1+i\over \textstyle 2}
\,\root\of{\sin\vartheta} \cr
\noalign{\smallskip}
{\textstyle 1-i\over \textstyle 2}
\,\root\of{\sin\vartheta} & \sin(\vartheta/2)
},
\label{IV-deg}
\ee
that transforms a two-dimensional vector at the lower vertical edge
to the one at the upper vertical edge, and each $S$-operator of the kagome
lattice does this independently.

The remarkable property of the matrix
(\ref{IV-deg}) is its {\em degeneracy}. Thus, the whole lattice
transfer matrix applied to the one-particle state yields
an {\em infinite sum of vectors proportional to a fixed vector}.
This sum must be inevitably equal to zero, due to the inf\/initeness
of lattice in all directions, the fact that a one-particle vector
just acquires some scalar factors under lattice translations,
and a logical assumption
\[
\sum_{n=-\infty}^{\infty} a^n =0
\]
for the sum of a geometrical progression inf\/inite in both directions.

\subsection{The marriage: an example of a string}
\label{sec-marriage}

The matrix (\ref{IV-deg}) is not the only degenerate one.
Formulae~(\ref{IV-A})--(\ref{IV-D}) show really that if we f\/ix
{\em any\/} values (0~or~1) for the four horizontal spins
of an $S$-operator, then the $2\times 2$-matrix corresponding to the
two remaining vertical edges is degenerate. Imagine now that we have f\/ixed
the spins at all horizontal edges of {\em all operators}. Then
edges with the spin~1 form ``strings'', and a remarkable conclusion
is that a conf\/iguration of those strings determines the ``outgoing''
vector at the upper vertical edges up to a scalar factor.

\medskip

\noindent
{\bf Remark.} However, the ``incoming'' vector at the lower edges
can cause this scalar factor to equal~0 for some string conf\/igurations.
For example, for a one-particle incoming vector, there can be no
closed strings, because they require at least two matrices~$D$
to be involved, as explained in Subsection~\ref{sec-absence}.

\medskip

We can propose the following
example of a string-like state resulting from the ``marriage''
of a one-particle state and a string within the horizontal plane.
The one-particle state is applied to the lower edges, as already
explained. It is supposed that the string is ``born'' in the vertex
where the particle is applied, due to the matrix~$D$ that can
change by~2 the total ``horizontal'' spin.
The form of the string at inf\/inity can be f\/ixed by two given
half-lines, each going along the edges of the lattice
in one of the east, north, or north-east directions.

\setcounter{equation}{0}
\section{Two cases of string superposition}
\label{V}

Let us recall that we have introduced in Section~\ref{I} some
``one-particle'' eigenstates for the model based upon solutions
of the tetrahedron equation. In the same Section, we have also
constructed some ``two-particle'' states. However, some
special condition arised in this construction, and the
superposition of two {\em arbitrary\/} one-particle states
was not achieved. Even the ``creation operators'' of Section~\ref{II}
did not give a clear answer concerning multi-particle states.

On the other hand, in Section~\ref{IV} we have brought in correspondence
to a one-particle state some new state that could be
called ``one open string''. It was done using some special ``kagome
transfer matrix''. Here we will show that the superposition
of such one-string states is easier to construct, because of
degeneracy of kagome transfer matrix: it turns into zero the
``obstacles'' that hampered constructing of multi-particle states.

The scheme of string --- particle ``marriage'' in Section~\ref{IV} was
as follows: take a one-particle state from Sections~\ref{I} and~\ref{II},
and apply
to it a kagome transfer matrix with boundary conditions corresponding
to the presence of two string tails in the inf\/inity, e.g. like this:
\
{\unitlength=1mm
\linethickness{0.4pt}
\begin{picture}(25.00,4.50)
\put(25.00,2.00){\oval(50.00,4.00)[l]}
\end{picture}\,}.

In this Section, we are going to complicate this scheme in the following
way: the boundary conditions will correspond to the presence
of an even number
of string tails at the inf\/inity, and instead of a one-particle
state, we will use some special multi-particle vector~$\Psi$.
Its peculiarity will be in the fact that $\Psi$ is {\em no longer
an eigenstate\/} of the hedgehog transfer matrix~$T$ def\/ined in
Section~\ref{I}. Instead, it will obey the condition
\be
T\Psi = \lambda\Psi + \Psi',
\label{V-int-1}
\ee
where $\lambda={\rm const}$, and $\Psi'$ is annulated by the
kagome transfer matrix of Section~\ref{IV}, which we will here
denote~$K$.

Recall that we have def\/ined $T$ in such a way that its degrees
could be described geometrically as ``oblique slices''
of the cubic lattice.
The transfer matrix~$T$ can be passed through the transfer matrix~$K$:
\be
TK=KT,
\label{V-int-2}
\ee
the boundary conditions (such as the number and form of tails
at the inf\/inity) for $K$ being intact. Def\/ine vector~$\Phi$ as
\[
\Phi=K\Psi.
\]
This together with (\ref{V-int-1}) and (\ref{V-int-2}) gives
\be
T\Phi=\lambda\Phi,
\label{V-int-4}
\ee
exactly as needed for an eigenvector.

\subsection{Eigenvectors of the ``several open strings'' type
for inf\/inite lattice}
\label{secV-1}

Let there be $n$ one-particle amplitudes $\varphi_{\ldots}^{(1)}, \ldots,
\varphi_{\ldots}^{(n)}$ of the same type as those described
in Section~\ref{I}. Let us compose an ``$n$-particle vector''
$\Psi$, i.e. put in correspondence to each unordered
$n$-tuple of vertices $A^{(1)},\ldots,A^{(n)}$ of the kagome lattice
some amplitude, in the following symmetrized way:
\be
\psi_{A^{(1)},\ldots,A^{(n)}} = \sum_s \varphi_{A^{s(1)}}^{(1)}
\ldots \varphi_{A^{s(n)}}^{(n)},
\label{V-1-1}
\ee
where $s$ runs through the group of all permutations
of the set $\{1,\ldots,n\}$.

As for the boundary conditions for the transfer matrix~$K$,
let us assume that there are exactly $2n$ string tails,
and they all go in positive directions,
that is lie asymptotically in the f\/irst quadrant.
Thus, in each of the points $A^{(1)},\ldots,A^{(n)}$ a string is
created, and they are not annihilated.

The vector (\ref{V-1-1}) is not an eigenvector of transfer matrix~$T$
due to problems arising when two or more points $A^{(k)}$ get close
to one another. Nevertheless, the vector $\Phi=K\Psi$ {\em is\/}
an eigenvector, because for it those problems disappear due to the
simple fact:
{\em creation of two or more strings within one triangle of the kagome
lattice is geometrically forbidden}.

\subsection{Eigenvectors of the ``closed string'' type for inf\/inite
lattice}
\label{secV-2}

In this Subsection, we will put in correspondence
to each unordered pair of vertices
of the inf\/inite kagome lattice an ``amplitude'' $\Psi_{AB}$
according to the following rules. If one of the vertices,
say $A$, {\em precedes\/} the other one, say $B$, in the sense
that they can be linked by a path --- a broken line --- going along
lattice edges in positive directions, namely northward, eastward,
or to the north-east, then let us put
\be
\Psi_{AB}= \varphi_A \psi_B - \psi_A \varphi_B,
\label{V-2-1}
\ee
where $\varphi_{\ldots}$ and $\psi_{\ldots}$ are two one-particle
amplitudes.
In the case if vertices $A$ and $B$ cannot be joined by a path
of such kind, let us put
\[
\Psi_{AB}=0.
\]

The values $\Psi_{AB}$ are components of the vector~$\Psi$
that belong to the two-particle subspace of tensor product of
two-dimensional spaces situated in all kagome lattice vertices.
What prevents $\Psi_{AB}$ from being an eigenvector of the hedgehog
transfer matrix is discrepancies arising near those pairs $A,B$
that lie at the ``border'' between such pairs where one of the
vertices precedes the other (so to speak, ``the interval $AB$
is timelike''),
and such pairs where it does not (``the interval $AB$
is spacelike'').

Those discrepancies, however, disappear for the vector $\Phi=K\Psi$,
where~$K$ is the kagome transfer matrix
with boundary conditions reading {\em no string tails at the
infinity}. This is because if a string cannot, geometrically,
be created at the point~$A$ (or~$B$) and then annihilated
at the point~$B$ (or~$A$), then the amplitude~$\Psi_{AB}$
doesn't inf\/luence at all the vector~$\Phi$.
The only thing that remains to be checked for (\ref{V-int-4})
to hold is a situation where $A$ and $B$ are in the same kagome lattice
triangle that will be turned inside out by one of the hedgehogs
of transfer matrix~$T$, as in Fig.~\ref{figV-1}.
Acting in the same manner as in Subsection~\ref{sec-twop}, write
\be
\pmatrix{ \varphi_{A'} \cr \varphi_{B'} }=
\pmatrix{\alpha & \beta \cr \gamma & \delta}
\pmatrix{ \varphi_A \cr \varphi_B }, \qquad
\pmatrix{ \psi_{A'} \cr \psi_{B'} }=
\pmatrix{\alpha & \beta \cr \gamma & \delta}
\pmatrix{ \psi_A \cr \psi_B },
\label{V-2-3}
\ee
where
\be
\alpha=-\delta, \qquad \alpha\delta-\beta\gamma=-1.
\label{V-2-4}
\ee
It follows from the formulas (\ref{V-2-3}) and (\ref{V-2-4}) that
\[
\varphi_A \psi_B-\varphi_B \psi_A=
\varphi_{B'} \psi_{A'}-\varphi_{A'} \psi_{B'},
\]
i.e.
\[
\Psi_{AB}=\Psi_{B'A'},
\]
exactly what was needed to comply with the fact that an
$S$-operator-hedgehog acts as a unity operator in the
two-particle subspace.
\begin{figure}[!ht]
\begin{center}
\unitlength=1mm
\linethickness{0.4pt}
\begin{picture}(65.00,33.00)
\put(3.00,12.00){\line(1,0){20.00}}
\put(23.00,12.00){\line(0,1){20.00}}
\put(23.00,32.00){\line(-1,-1){20.00}}
\put(30.00,17.00){\vector(1,0){6.00}}
\put(43.00,2.00){\line(0,1){20.00}}
\put(43.00,22.00){\line(1,0){20.00}}
\put(63.00,22.00){\line(-1,-1){20.00}}
\put(3.00,12.00){\circle*{1.00}}
\put(23.00,12.00){\circle*{1.00}}
\put(23.00,32.00){\circle*{1.00}}
\put(43.00,22.00){\circle*{1.00}}
\put(63.00,22.00){\circle*{1.00}}
\put(43.00,2.00){\circle*{1.00}}
\put(13.00,22.00){\vector(1,1){1.00}}
\put(23.00,22.00){\vector(0,1){1.00}}
\put(13.00,12.00){\vector(1,0){1.00}}
\put(43.00,12.00){\vector(0,1){1.00}}
\put(53.00,12.00){\vector(1,1){1.00}}
\put(53.00,22.00){\vector(1,0){1.00}}
\put(1.00,12.00){\makebox(0,0)[rc]{$A$}}
\put(25.00,32.00){\makebox(0,0)[lc]{$B$}}
\put(41.00,2.00){\makebox(0,0)[rc]{$B'$}}
\put(65.00,22.00){\makebox(0,0)[lc]{$A'$}}
\end{picture}
\end{center}

\vspace{-6mm}

\caption{}
\label{figV-1}
\end{figure}

\setcounter{equation}{0}
\section{Discussion}

In this paper, we were not trying to discuss the ``physical''
consequences of the eigenvectors constructed. Its only modest aim
was to show that some ideas of the classical Bethe ansatz could be relevant
for $2+1$-dimensional models and that, on the other hand, there arise
new structures, inherent for multidimensions, namely
``string-like'' eigenvectors.

The $S$-operator on which our model is based was discovered by the author
in 1989~\cite{k1}.
Later on, a similar but dif\/ferent model was discovered
by J.~Hietarinta~\cite{h}, and then it was shown
by S.M.~Sergeev, V.V.~Mangazeev and Yu.G.~Stroganov~\cite{mss} that
both those models are particular cases of one model, parallel
in some sense to the Zamolodchikov model. This allows one to hope
that the rich mathematical structures already discovered in
connection with our model will be extended some time onto
the Zamolodchikov model as well.

As we were only able to construct one- and two-particle Bethe-like vectors
Section~\ref{I}, it was quite natural to try to do more with
the help of ``algebraization'' of their construction in Section~\ref{II}.
However, here again too big complications arise when trying to actually
construct more eigenvectors.
In particular, even the superposition of two {\em arbitrary\/}
one-particle states from Section~\ref{I} has not been obtained yet.

It is clear that the superposition of two arbitrary one-particle
eigenstates from Section~\ref{I} cannot lie just in the 2-particle
space as it is def\/ined in Section~\ref{I}. So, probably, some new
terminology should be introduced to distinguish between the
2-{\em particle\/} space and 2-{\em excitation\/} states.

The Sections~\ref{I}, \ref{II} and~\ref{III} together show that
\begin{itemize}
\item
even the model corresponding to the simplest solutions of tetrahedron
equation possesses a large variety of eigenstates which are probably
not easy to classify,
\item
eigenvalues seem sometimes rather trivial --- maybe, it is due to
relation $S^2={\bf 1}$, see~\cite{k2} --- but probably they will be
more interesting for the models from \cite{h,mss},
\item
some states can be introduced that are countable sums of
formal tensor products
throughout the inf\/inite lattice, but it is unclear what to do
for a f\/inite lattice,
\item
there probably does not exist --- at least, it was not
discovered in papers~\cite{k1,k2,h,mss} --- a complete analog
of the 6-vertex model in $1+1$ dimensions with its
``conservation of particle number'', but something can
be built even upon the fact that that number is conserved ``sometimes'',
\item
there exists a huge amount of symmetries multiplying eigenvalues
by constants (roots of unity for a f\/inite lattice) and
unknown for the $1+1$-dimensional models, and
\item
eigenstates can be constructed with making no use of {\em invertibility\/}
of tetrahedron equation so\-lu\-tions --- so probably it makes sense
to search for non-invertible ones.
\end{itemize}

In Sections \ref{IV} and~\ref{V} we provide what seems to be
the most promising way of constructing new eigenstates.
Those are, in essence,
strings from Section~\ref{III} fertilized by nontrivial momentum particles
of Sections~\ref{I} and~\ref{II}. Such strings
are not just invariant under the shifts
along themselves, but acquire some nontrivial multipliers.

It is clear that really a quite immense zoo
of such states can be constructed.

The main point is, however, that the string --- particle ``marriage''
makes possible a simple and clear construction
of at least some multi-string states. At least some obstacles at which
we have run into when trying to construct multi-{\em particle\/} states
miraculously disappear when we add a string to each particle.

The mentioned states have been constructed for the inf\/inite kagome lattice.
We have to recognize that constructing such states on a f\/inite
lattice remains an open problem.

It is also unclear how to combine the results
of Subsections \ref{secV-1} and~\ref{secV-2}, i.e. construct
such states with string ``creation'' and ``annihilation'' where
the total number of ``creating'' and ``annihilating'' particles
would be more than two. Note that in Subsection~\ref{secV-1} we use
the symmetrized product of one-particle amplitudes, while
in Subsection~\ref{secV-2} --- the antisymmetrized one.

Concerning the dispersion relations of Subsection~\ref{V-sec-disp},
let us remark that perhaps there are too many of them.
It is probably caused by the fact that, for now, we managed
to construct not all one-particle and/or one-string states.

On the other hand,
it is clear that the dispersion relations of type
(\ref{V-3-6})--(\ref{V-3-7}) survive also for a string ``created
by a particle''. As for the multi-string states,
all of the eigenvalues are obtained for them as products
of corresponding eigenvalues for each string.

Finally, it is certainly interesting to f\/ind eigenstates for the model
based on other simple solutions to the tetrahedron equation~\cite{h},
and perhaps for the general model described in~\cite{mss}.

\subsection*{Acknoledgements}

 I am grateful to H.~Boos, R.~Kashaev, V.~Mangazeev,
S.~Sergeev and Yu.~Stroganov for useful comments on the preprint
version~\cite{5 preprintov} of this paper.
I also acknowledge the Russian Foundation for Basic Research
under Grant no.  98-01-00895.

\label{korepanov-lp}

\end{document}